\documentclass[5p,sort&compress,times,fleqn]{elsarticle}

\usepackage{graphicx}
\usepackage{wrapfig}
\usepackage{caption}
\usepackage{comment}
\usepackage{hyperref}
\usepackage{subfig}

\usepackage[utf8x]{inputenc}
\usepackage{amsmath}
\usepackage{amssymb}
\usepackage{graphicx}
\usepackage[normalem]{ulem}
\usepackage{epsf,epsfig}
\usepackage{psfrag}
\usepackage{comment}

\newcommand{\be}{\begin{equation}}
\newcommand{\ee}{\end{equation}}
\newcommand{\bea}{\begin{eqnarray}}
\newcommand{\eea}{\end{eqnarray}}

\def\mpl{m_{\rm Pl}}

\begin{document}

\title{Constraining the loop quantum gravity parameter space from phenomenology}

\author{Suddhasattwa Brahma}
\ead{suddhasattwa.brahma@gmail.com}
\address{Center for Field Theory and Particle Physics, Fudan University,
	Shanghai 200433, China}
\address{Asia Pacific Center for Theoretical Physics, Pohang 37673, Korea}

\author{Michele Ronco}
\ead{michele.ronco@roma1.infn.it}
\address{Dipartimento di Fisica, Universit\`a di Roma ``La Sapienza", P.le A. Moro 2, 00185 Roma, Italy}
\address{INFN, Sez.~Roma1, P.le A. Moro 2, 00185 Roma, Italy}

\begin{abstract}
Development of quantum gravity theories rarely takes inputs from experimental physics. In this letter, we take a small step towards correcting this by establishing a paradigm for incorporating putative quantum corrections, arising from canonical quantum gravity (QG) theories, in deriving \textit{falsifiable} modified dispersion relations (MDRs) for particles on a deformed Minkowski space-time. This allows us to differentiate and, hopefully, pick between several quantization choices via \textit{testable, state-of-the-art} phenomenological predictions. Although a few explicit examples from loop quantum gravity (LQG) (such as the regularization scheme used or the representation of the gauge group) are shown here to establish the claim, our framework is more general and is capable of addressing other quantization ambiguities within LQG and also those arising from other similar QG approaches.
\end{abstract}

\maketitle


\section{Introduction and motivation} 
\noindent
It is well-known that the lack of experimental evidence represents one of the main obstacles in our search for a theory of QG \cite{oriti,smol,smam,gacLRR}. In the absence of observations, researchers often rely on less dependable principles, such as `beauty' and `naturalness', as guidance for advancing QG proposals \cite{oriti1}. Working within a given approach, one is then usually forced to choose between quantization ambiguities, often on the same footing theoretically, by following one's personal penchants or other questionable criteria. In this letter, we relate different quantization schemes, which have been proposed in the LQG literature  \cite{Amb1,alex,Amb2,Amb3}, to different predictions for observable quantities. And by doing so, we lay down a framework to distinguish between them using observations.

LQG is a non-perturbative, background-independent approach to quantize gravity \cite{RovLRR, AshLew}, with significant accomplishments such as `singularity resolution' in various cosmological and black-hole scenarios \cite{BojoLRR,LQGbh}. However, as in other QG models, conclusions typically depend on various quantization choices. So far very little work has been directed towards understanding whether these formal alternatives affect physical outcomes. Among other reasons, this is largely a consequence of the fact that the complexity of the full-fledged theory has created a gap between technical results and potential observations.  

Remarkably, recent results in symmetry-reduced LQG models which, in particular, has focussed on the study of quantum symmetries in the presence of LQG-inspired corrections \cite{SC1,SC2,oper,bojo2,bojo3}, have unanimously discovered the  fact that general covariance should be \textit{modified} by such quantum effects. These modifications amount to a deformation of the brackets closed by the gravitational constraints which generate space and time gauge transformations. Here, we outline a path to derive MDRs  corresponding to the modified brackets and show that quantization ambiguities leave their imprints on the form of the MDR. This would suggest that different quantization schemes adopted (and often treated interchangeably) are not equivalent and, conceivably, might be distinguished thanks to forthcoming tests of Planck-scale departures from special relativistic symmetries \cite{gacLRR,mattLRR,alan,sabine,barrau}.

Although we focus on particular quantization choices characteristic to LQG (such as the choice of the Immirzi parameter, the regularization scheme used or the dimension of the gauge group), we shall unequivocally demonstrate that our analysis is general enough to include other such ambiguities in LQG as well as for corrections coming from other canonical QG approaches.  

Throughout the paper we work with natural units if not specified differently.

\section{Deformed covariance and modified dispersion relations}
\noindent
One of the newest results in LQG is the emergence of non-classical space-times structures \cite{bojoPa1, bojoSR,vara,tom,pranper,kowros,norb,bojo1,lqgbook}. These departures from smooth classical space-time manifolds can be meaningfully traced back to quantum modifications of the so-called hypersurface deformation algebra (HDA), which encodes covariance in the Hamiltonian formulation of classical general relativity \cite{adm}. In fact, since the structure function appearing in the classical HDA is the inverse of the spatial metric $h^{ab}$ on the hypersurface (see e.g. \cite{bojobook} as well as the equation below), then it is believed that any modification (usually by a phase-space function, $\beta$) to it points towards a deformation of the space-time geometry. The common feature of LQG models is that only the bracket between two generators of normal deformations (or the Hamiltonian constraint) is affected:
\begin{equation}
\label{defHDA}
\{ H^Q[N], H^Q[M] \} = D[\beta h^{ab}(N\partial_b M -M\partial_b N)] \,.
\end{equation}
Gauge transformations generated by constraints ($H$\footnote{The superscript `Q' implies that we are dealing with the LQG quantum-corrected Hamiltonian constraint here.} and $D$ for normal $N$ lapse and tangential $N^a$ shifts respectively) represent coordinate freedom in classical canonical gravity. The closure of the brackets assures there is no violation of the gauge symmetries, but the modification in the above equation implies a \textit{deformed} notion of covariance \cite{bojo1, bojo2, bojo3}. 

It is known that the Poincar\'e algebra, which describes symmetries of Minkowski space-time, can be derived as a special case from the classical HDA in a systematic manner \cite{bojobook,regge} (see below for a short review of the procedure). Not surprisingly, LQG-deformations of the HDA turn into corresponding deformations of the Poincar\'e algebra \cite{defPaLQG,defPaLQG1,ncLQG}. As a consequence, as first shown in \cite{phenoLQG}, the familiar dispersion relation (for massless particles), $E^2 = p^2$,  does not hold true anymore and is replaced by more complicated expressions. 

Usually, due to the complexity of QG theories, MDRs are either parametrized as a generic series expansion in (inverse) powers of $\mpl$ with some unknown coefficients, i.e. $E^2 \simeq p^2 + a_1 p^3/ \mpl + a_2 p^4 / \mpl^2$ (where $a_1, a_2, \dots , a_n$ are to be determined experimentally) as a purely phenomenological ansatz, or derived in simplified models (see e.g. \cite{gacLRR,dsr,MagSmo,majRue,Lukrue}). Taking the opposite direction here, we compute MDRs from a fundamental QG theory -- LQG -- and, thus, contribute to bridge the gap between top-down and bottom-up approaches. From this perspective, our work is also part of an ongoing effort \cite{ncLQG,polyLQG} aimed at characterizing the Minkowski limit of LQG, and exploring if there is any relation to non-commutative geometries \cite{dfr,ws,MajAm,posop} as a way to characterize the so-called \textit{spacetime fuzziness} or \textit{foaminess} \cite{fuzz1,fuzz2,fuzz3,fuzz4}.

\subsection{Deriving MDRs from deformed-HDA}
\noindent
In this section, we take the Minkowski limit of the LQG-deformed HDA and eventually prove that it affects the dispersion relation through a corresponding deformation of the Poincaré algebra.  The full deformed-HDA is given by 
\begin{equation}\label{qhda}
\begin{split}
\{D[M^{a}],D[N^{a}]\}=D[\mathcal{L}_{\vec{M}}N^{a}],\\
\{D[N^{a}],H^{Q}[M]\}=H^{Q}[\mathcal{L}_{\vec{N}}M],\\
\{H^{Q}[M],H^{Q}[N]\}=D[\beta h^{ab}(M\partial_{b}N-N\partial_{b}M)] \, .
\end{split}
\end{equation}
In order to reduce to the flat limit, one has to restrict to linear lapse and shift functions, which correspond to linear coordinate changes, i.e.
\begin{equation}
\label{minklim}
N^{k}(x) = \Delta x^{k}+R^{k}_{i}x^{i}\quad N(x) = \Delta t+v_{i}x^{i}
\end{equation}
and, at the same time, to flat spatial hypersurfaces i.e. $h_{ij} \equiv \delta_{ij}$.  With these restrictions one can prove that the infnite set of general diffeomorphisms reduce to the finite subset of Poincaré transformations. It is then possible to read off the commutators between the Poincar\'e generators directly from the HDA.In particular, from $\{D[M^{a}],D[N^{a}]\}$ one can derive $\{J_i,J_j\}, \{J_i,P_j\}$, and $ \{P_i,P_j\}$ ($J_i$ being the generator of rotations and $P_i$ that of spatial translations), while  $\{J_i,N_j\}, \{P_0,J_j\},  \{N_i,P_j\}$ and $\{P_i,P_0\}$ ($N_i$ being the generator of boosts and $P_0$ that of time translations) can be obtained from $\{D[N^{a}],H^{Q}[M]\} $, and finally from $\{H^{Q}[M],H^{Q}[N]\}$ one gets $\{N_i,N_j\}$ and $\{N_i,P_0\}$. In the appendix, we explicitly illustrate the case of rotations as an example.  

Let us start with the spherically-symmetric reduction of Hamiltonian gravity in Ashtekar-Barbero variables (see e.g.\cite{olmedo}) in the presence of LQG deformations. In this case the ADM foliation ~\cite{adm} allows to decompose the space-time manifold as $\mathcal{M} = \mathbb{R}\times \Sigma = \mathcal{M}_{1+1}\times S^2$, where $\mathcal{M}_{1+1}$ is a 2-dimensional manifold spanned by $(t,r)$ and $S^2$ stands for the 2-sphere. Given that, the line element reads
\begin{equation}
\label{metric}
ds^2 = -N^2dt^2+h_{rr}(dr+N^r dt)^2 + h_{\theta\theta}(d\theta^2+\sin^2\theta\varphi^2)\,,
\end{equation}
where the shift vector is purely radial, {\it i.e.} $N^i = (N^r,0,0)$, due to spherical symmetry, and, consequently, we are left only with radial diffeomorphisms generated by $D[N^r] = \int dr N^r \mathcal{H}_r$ (where $\mathcal{H}_r$ is the only non-vanishing component of the momentum density) and, time transformations, generated by $H[N] = \int dr N \mathcal{H}$ (where $\mathcal{H}$ is the Hamiltonian density). The components of the spatial metric $(h_{rr},h_{\theta\theta})$ can be written in terms of rotationally invariant densitized triads which are given by:
\begin{eqnarray}\label{triads}
E&=&E^{a}_{i}\tau^{i}\frac{\partial}{\partial x^{a}}=E^{r}(r)\tau_{3}\sin\theta\frac{\partial}{\partial r}+\nonumber\\
&&\hspace{1cm}+E^{\varphi}(r)\tau_{1}\sin\theta\frac{\partial}{\partial\theta}+E^{\varphi}(r)\tau_{2}\frac{\partial}{\partial\varphi}\,,
\end{eqnarray}
where $\tau_{j}=-\frac{1}{2}i\sigma_{j}$ represent \textit{SU}(2) generators. The densitized triads are canonically conjugate to the extrinsic curvature components, which, in presence of spherical symmetry, are conveniently
described as follows
\begin{equation}
\begin{split}
K=K^{i}_{a}\tau_{i}dx^{a}=K_{r}(r)\tau_{3}dr+K_{\varphi}(r)\tau_{1}d\theta+\\+K_{\varphi}(r)\tau_{2}\sin\theta d\varphi\,.
\end{split}
\end{equation}
For the simplest case including only local holonomy corrections \cite{ed,fra}, with $\gamma \in \mathbb{R}$ and $j=1/2$, the deformation $\beta$ takes the form
\begin{equation}
\beta=\cos(2\delta K_{\varphi}) \, , 
\end{equation}
where $\delta$ is a regularization parameter,  related to the square root of the minimum eigenvalue of the area operator.

As already discussed in \cite{ncLQG}, the main difficulty lies in the fact that LQG-deformations in the HDA arises in the form of the structure function getting modified by a function of the phase space variables, while deformations at the level of the Poincar\'e algebra usually implies modification of the algebra generators \cite{majRue,Lukrue}. As a way out, it is then convenient to find a way to write $\beta$ in terms of symmetry generators (see also \cite{defPaLQG,defPaLQG1,phenoLQG}), and for this purpose, it is valuable to notice that observables of the Brown York momentum~\cite{BY},
\begin{equation}\label{bymom}
P=2\int_{\partial\Sigma} d^{2}z\upsilon_{b}(n_{a}\pi^{ab}-\overline{n}_{a}\overline{\pi}^{ab}) \, ,
\end{equation}
can be identified by extrinsic curvature components provided one makes a suitable choice for $\delta \propto |E^{r}|^{-\frac{1}{2}}$. In Eq.\eqref{bymom}, we have that $\upsilon_{a}=\partial/\partial x^{a}$, $n_{a}$ is the co-normal of the boundary of the spatial region $\Sigma$, and $\pi^{ab}$ plays the role of the gravitational momentum (while the over barred symbols in the above equation are the same functions but evaluated at the boundary). From this, it is possible to establish that the radial Brown-York momentum $P_{r}$ is related to the extrinsic curvature component $K_{\varphi}$ in the following way 
\begin{equation}\label{bymomf}
P_{r}=-\frac{K_{\varphi}}{\sqrt{|E^{r}|}}.
\end{equation}
Thus, we also have that 
\begin{equation}\label{funb}
\beta = \cos(\lambda P_{r}) \, , 
\end{equation}
where $\lambda$ is a parameter of the order of the Planck length ($\lambda \sim 1/ \mpl$\footnote{Keep in mind that its exact value also depends on quantization ambiguities.}). 

At this point we can implement the Minkowski limit by taking $N = \Delta t + v_{r}r$ and $N^r = \Delta r$ in Eq.\eqref{minklim}, and one finds from Eq.\eqref{qhda}
that it is characterized by a deformed commutator
\begin{equation}\label{bjcommut}
[B_{r}, P_{0}] = iP_{r}\cos(\lambda P_{r}) \, .
\end{equation}
Since only the Poisson bracket involving two scalar constraints  is quantum corrected (see Eqs. \eqref{qhda}), the other commutators are undeformed, i.e. $[B_{r}, P_{r}] = iP_{0}$ and $[P_{0}, P_{r}] = 0$. This derivation shows how the modifications of the HDA translate into a corresponding departure from the standard special-relativistic symmetry structure. Planck-scale deformations of special relativity have been already considered by many authors (see e.g. \cite{dsr,MagSmo} for Doubly Special Relativity proposals) within the framework of QG and some of the most interesting results admit a mathematical formulation in terms of Hopf algebras or quantum groups \cite{majRue,Lukrue}. A first attempt to interpret Eq. (\ref{bjcommut}) as a Hopf algebraic structure has been done by one of us in \cite{ncLQG}. The interested reader can also take a look at \cite{polyLQG} for a different discussion along the same lines and drawing similar conclusions. A major obstruction which prevents from reaching a conclusive result is the lack of information on the coalgebra structure, which is usually the trademark of non-trivial modifications of standard Lie algebras. However, independent of its relation with quantum groups, the above commutator can be written in terms of ADM charges which are directly connected to boundary observables or stress-energy components \cite{adm},  as noticed already in  \cite{bojoPa1}. Thus, one should not be allowed to arbitrarily choose the generators of the algebra \cite{bojoPa1}. What is more, in light of this, the authors of \cite{bojoPa1} suggest that the derivation of Poincar\'e deformations from the HDA may contribute to solve the issue regarding the \textit{choice of basis} on momentum space which represents an element of ambiguity in the current literature.

For our purposes, it is important to see that this symmetry deformation implies a modification of the mass Casimir (for massless particles) of the form (as an be verified by a straightforward check by the reader)
\begin{equation}
P^2_0 = \int \beta(P_r) P_r dP_r  \, , 
\end{equation}
whose explicit expression depends on the particular corrections implemented. For instance if $\beta$ is given by Eq. \eqref{funb} then we obtain 
\begin{equation}
P^2_0 = -2\lambda^{-2} + 2\lambda^{-2}(\cos( \lambda P_r) + \lambda P_r \sin(\lambda P_r))
\end{equation}
and, thus, upon the identifying $P_0 \sim E$ and $P_r \sim p$, we find the modified on-shell relation $E^2 =   -2\lambda^{-2} + 2\lambda^{-2}(\cos( \lambda p) + \lambda p \sin(\lambda p))$. In the next section we give the MDR for different deformation functions $\beta$ which have been motivated by recent LQG-based analyses appeared in the literature. 

\subsection{Immirzi parameter} 
\begin{figure}[!tbp]
  \centering
  \subfloat[]{\includegraphics[width=0.4\textwidth]{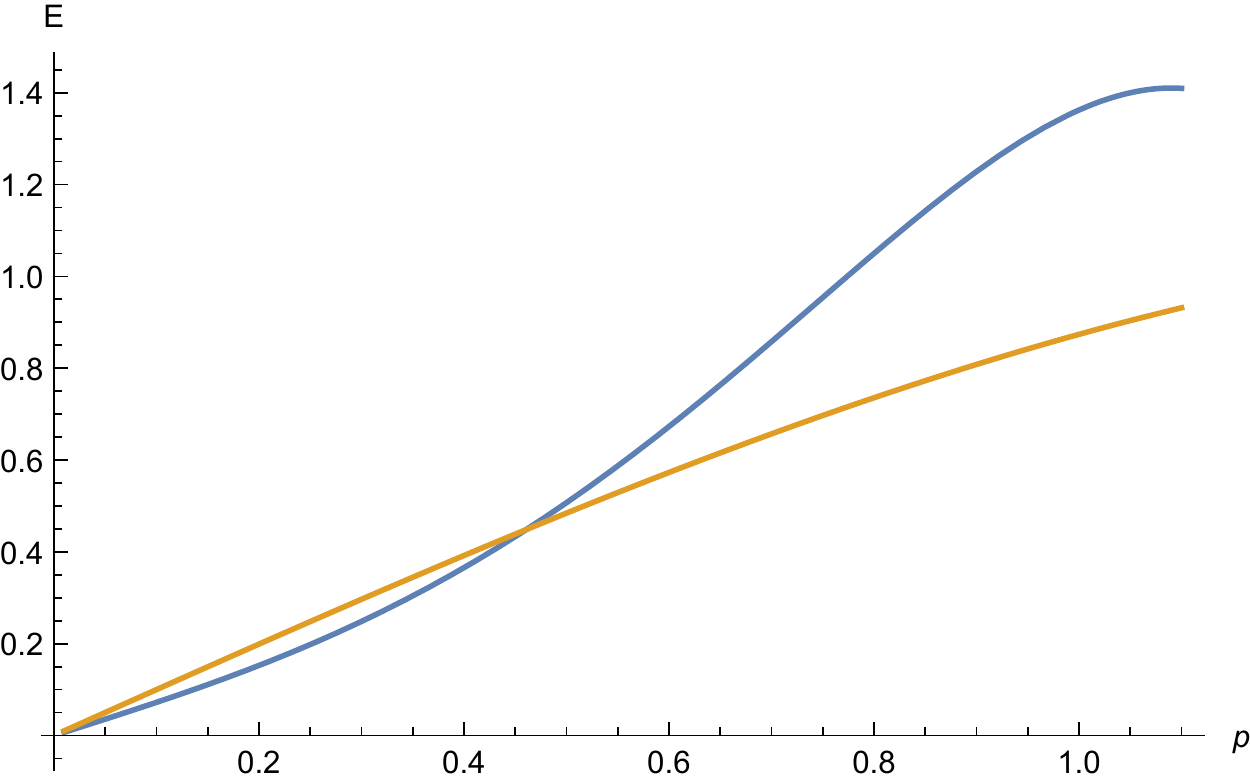}\label{fig:sub1}}
  \hfill
  \subfloat[]{\includegraphics[width=0.4\textwidth]{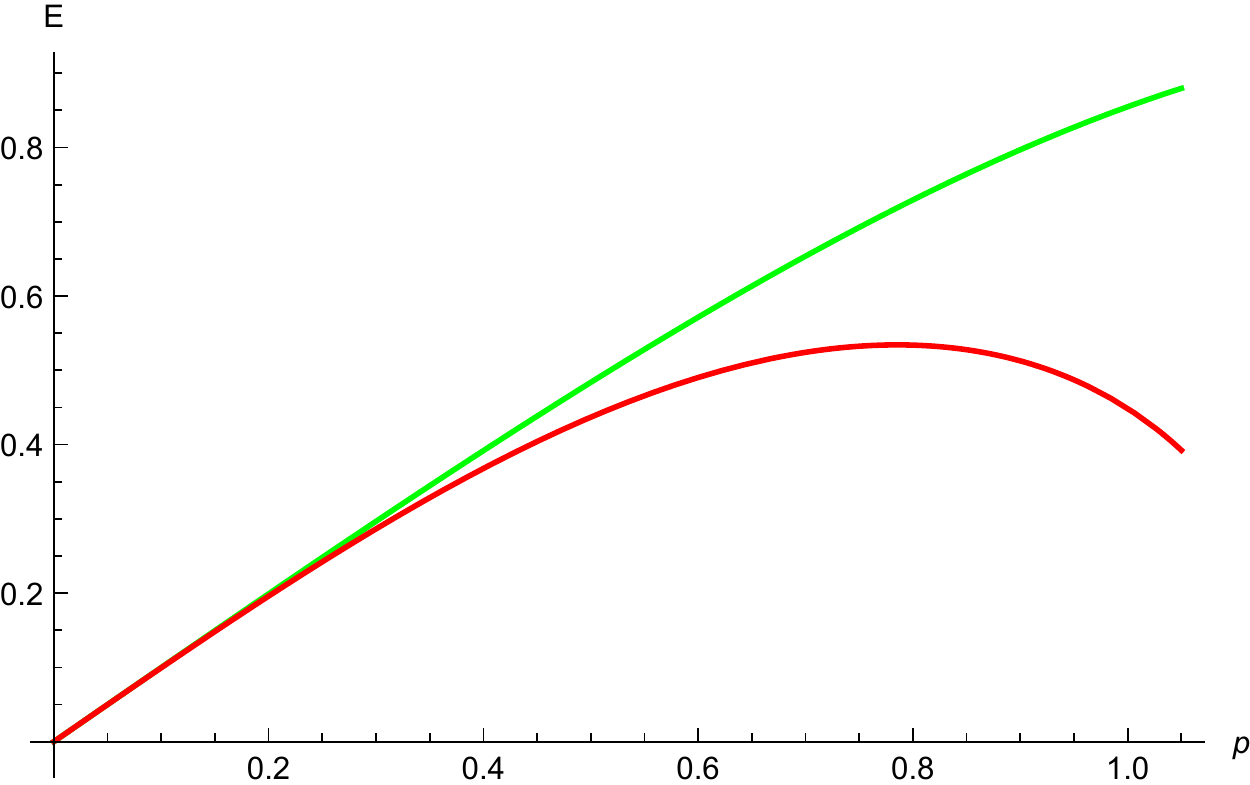}\label{fig:sub2}}
  \caption{The graphs compares different MDRs obtained for LQG holonomy-corrections, within different quantization schemes: on the top (a),  the orange plot represents the $SU(2)$ case with real $\gamma$ while the blue plot corresponds to the choice of a purely imaginary $\gamma$ implemented in the $SU(1,1)$ gauge. They are both in the fundamental representation. On the bottom (b) we have two MDRs, both calculated for a real $\gamma$ and  in the $j=1$ representation, but using two different methods of regularizing the field strength: the green plot for the `holonomy' regularization and the red plot for the `connection'  regularization. The orange plot in (a) and either of the plots in (b) compares MDRs for different spin values, $1/2$ and $1$ respectively. We set $\mpl \equiv 1$.}
\end{figure}

\noindent
As explicitly shown above, the resulting MDR is determined by $\beta$ which, in turn, depends on quantization choices within LQG as is crucial to our goal. For instance, a real or complex-valued Barbero-Immirzi parameter $\gamma$ \cite{barbero,immirzi}, a free parameter within LQG, generates quantitatively different MDRs \cite{phenoLQG}. Here we give the leading order correction to the dispersion relation for a real-valued $\gamma$ (set to $1$) and an imaginary $\gamma$ (with a $SU(1,1)$ gauge choice \cite{su11})\footnote{Both these calculations have been done for the  fundamental spin representation.}, which can be calculated from the modified mass-Casimir due to the holonomy-corrected HDA \cite{phenoLQG}. (The full expressions for the MDRs are compared in Fig. (\ref{fig:sub1}).) 
\bea
E^2 & \simeq &  \, p^2 -\frac{p^4}{4\mpl^2} + \mathcal{O}\left(1/\mpl^4\right)\, \;\;\; \gamma\in\mathbb{R}, \label{su2}\\
E^2 & \simeq & p^2 +\frac{p^4}{4\mpl^2} + \mathcal{O}\left(1/\mpl^4\right)\, \,\;\;\; \gamma = i.\label{su11}
\eea
Thus, if we were able to distinguish the sign (or even more optimistically the full coefficient) of the leading correction from experimental data, then it would be possible to single out a preferred choice for $\gamma$ by using observations. In particular, it is easy to realize that the former equation would allow subluminal motion v$\simeq 1-3E^2/(8\, \mpl)^2 $, while according to the latter MDR one would have v$\simeq 1 +3E^2/(8\, \mpl)^2$ implying superluminal motion. In the next paragraph, we show that similar considerations apply also to other formal choices such as the regularization scheme adopted to define quantum operators. 

\subsection{Regularization schemes}
\noindent
Even for a real-valued $\gamma$, conclusions still depend on other quantization ambiguities. Holonomy-corrections in LQG arise from regularizing the curvature operator in terms of holonomies of connections instead of the connections themselves. There are two main ways, which are somewhat misleadingly called the the `holonomy' (HR) and `connection' regularizations (CR) \cite{reg1,reg}, in which one can carry this out. In the former case, one uses the holonomy of a square plaquette to regularize the curvature operator while in the latter case, one uses open holonomies for achieving it. 

Similarly, the dimension (or spin) of the representation also plays a crucial role in the regularization procedure. Although there are sometimes justifications provided for using the fundamental representation in symmetry-reduced models in the form of choosing highly fine-grained states by packing them with a collection of units carrying the smallest quanta of geometry \cite{LQCreview}, this is somehow in contrast to the full theory where the states depend on different spin-labels. The spin-ambiguity in LQG also affects dynamics as the Hamiltonian constraint operator depends on its choice \cite{reg, spinam, spin2}.

Without going into details, it is important to point out that these are quantization ambiguities which are peculiar to LQG and also have impact on the dynamics of the theory. Following the procedure we explained above, we derived MDRs (see next section for the full expressions) for different spin-representations $j$ \cite{Amb2,spinam, spin2} as well as for the two different curvature-regularization schemes which exist in the literature. In Fig. (\ref{fig:sub2}), as an illustrative example, we compare MDRs for $j=1$ representation for the these two different regularization schemes (the two regularization schemes match only for the fundamental spin-$1/2$ representation but disagree for higher spins). Their leading-order expansions are
\bea
\label{reg2}
E^2 &\simeq& \,  p^2  -\frac{p^4}{4\mpl^2} + \mathcal{O}\left(1/\mpl^4\right) \, , \hspace{.8cm} \text{HR}.\\
E^2 &\simeq& \,  p^2  -\frac{p^4}{\mpl^2} + \mathcal{O}\left(1/\mpl^4\right) \, , \hspace{1cm} \text{CR}.
\eea
The common conclusion of these LQG corrections we obtain is that they are suppressed by two powers of the Planck mass, thereby being beyond our present experimental prowess but possibly detectable in the not-so-distant future (see e.g. \cite{gacLRR} and \cite{mattLRR} for a discussion on the perspectives of QG phenomenology and some encouraging estimates in astrophysical contexts). A well-known and extensively explored way to constrain MDRs comprises looking at the effect of in-vacuo dispersion \cite{grb1}, a direct consequence of energy-dependent modifications of particles' dispersion relations, with both photons \cite{magic,glast,veritas,fermi1,fermi2,fiore,nava,data2,data3,data4} and neutrino data \cite{data1,data5,icecube,km3}, and, perhaps, also combined soon with gravitational wave measurements \cite{gw,gw2}. It is not unconceivable that we may be able to explore $\mathcal{O}(1/\mpl^2)$ effects by combining three pivotal complementary factors: the collection of larger samples, the improvement of experimental sensitivities, and multi-messenger detections. 

Once we have access to these terms, as shown by our analysis, it might be possible to discriminate between such choices from the form of the MDR. Then, our results may provide a criterion for settling a wide class of quantization ambiguities. In fact, we stress that all the apparatus laid out in this work is applicable for \textit{any} deformation function $\beta$ arising from another QG approach, and not just the ones appearing in LQG, as long as the deformation appears only in the quantum-corrected Hamiltonian operators (as in \eqref{defHDA}). Of course, for LQG, this method is easily generalized to other quantization ambiguities not considered here. While the derivation of full expressions for MDRs corresponding to different choices of the Immirzi parameter has been addressed in \cite{phenoLQG}, those for different regularization choices and spin-representations are summarized in the next section.

\subsection{Full expression of MDRs for regularization schemes}
\noindent
The deformation functions for the HR scheme are listed  below corresponding to different spin-representations. 
\begin{enumerate}
	\item $\beta_{\frac{1}{2}} = \cos(2\delta K_{\phi})$ for holonomies calculated in the $j = 1/2$ representation;
	\item $ \beta_1 = \cos^3(\delta K_{\phi}) - \sin^4(\delta K_{\phi}) - \frac{7}{4} \sin(\delta K_{\phi}) \sin(2\delta K_{\phi})$\\ $+  \frac{3}{4} \sin(2\delta K_{\phi})^2$ for holonomies calculated in the $j = 1$ representation;
	\item $\beta_{\frac{3}{2}} =  -\sin^2(\delta K_{\phi}) + \frac{12}{5} \sin(\delta K_{\phi})^4 - \frac{9}{10} \sin(\delta K_{\phi})^6 + 
	\cos(\delta K_{\phi})^2 (1 +\frac{9}{2} \sin(\delta K_{\phi})^4) - \frac{39}{10} \sin(\delta K_{\phi})^3 \sin(2\delta K_{\phi}) + 
	\sin(2\delta K_{\phi})^2 (-\frac{9}{5} + \frac{9}{10} \csc(\delta K_{\phi}) \sin(2\delta K_{\phi}))  $ for holonomies calculated in the $j = \frac{3}{2}$ representation.
\end{enumerate}

Taking the Minkowski limit we can find a correspondingly deformed Poincar\'e algebra and, then, a modification of the energy-momentum dispersion relation. Explicitly, for massless particles: 
\begin{enumerate}
	\item $E^2 = -2 + 2(\cos( p) + p \sin(p))$ for spin $j = 1/2$;
	\item $E^2 = -\frac{3}{4} -  \cos(\frac{1}{2}p) +  \cos( p) + \cos(\frac{3}{2} p) - \frac{1}{4} \cos(2 p) - 
	\frac{p}{2} \sin(\frac{p}{2}) + p \sin(p) + \frac{3}{2} p \sin(\frac{3}{2} p) - \frac{p}{2} \sin(2 p)$ for spin $j = 1$;
	\item $E^2 = -\frac{23}{40} - \frac{3}{5} \cos(\frac{p}{2}) + \frac{13}{80} \cos(p) + \frac{9}{10} \cos(\frac{3p}{2}) +  \frac{3}{8} \cos(2p) - \frac{3}{10} \cos(\frac{5p}{2}) + \frac{3}{80} \cos(3p) -  \frac{3}{10} p \sin(\frac{p}{2}) + \frac{13}{80} p \sin(p) + \frac{27}{20} p \sin(\frac{3p}{2}) +  \frac{3p}{4}  \sin(2p) - \frac{3p}{4} p \sin(\frac{5p}{2}) + \frac{9}{80} p \sin(3p)$ for spin $j = 3/2$.
\end{enumerate}

The $\beta$ for different spin-representations of the CR scheme are as follows
\begin{enumerate}
	\item $\beta_{\frac{1}{2}} =  \cos(2\delta K_{\phi})$ for holonomies calculated in the $j = 1/2$ representation;
	\item $\beta_{1} =  \cos^4(\delta K_{\phi}) + \sin(\delta K_{\phi})^4 - \frac{3}{2} \sin(2\delta K_{\phi})^2$ for holonomies calculated in the $j = 1$ representation;
	\item $\beta_{\frac{3}{2}} =\sin^2(\delta K_{\phi}) + \frac{24}{5}\sin^4(\delta K_{\phi}) - \frac{18}{5} \sin^6(\delta K_{\phi}) + 
	\cos^2(\delta K_{\phi}) (1 + 18 \sin^4(\delta K_{\phi})) - \frac{18}{5}\sin^2(2\delta K_{\phi})$ for holonomies calculated in the $j = \frac{3}{2}$ representation.
\end{enumerate}

The corresponding (unexpanded) forms of the MDRs, for the above-mentioned $\beta$s, are given below. 
\begin{enumerate}
	\item $E^2 = -2 + 2(\cos(p) + p \sin( p))$ for spin $j = 1/2$;
	\item $E^2 =-\frac{1}{2} + 2 (\frac{1}{4} \cos(2 p) +  \frac{1}{2} p \sin(2p))   $ for spin $j = 1$;
	\item  $E^2 = \frac{1}{10} + 2\times 10^{-16} p^2 - \frac{11}{20} \cos(p) + \frac{3}{10} \cos(2p) + 
	\frac{3}{20} \cos(3p) - \frac{11}{20} p \sin(p) + \frac{3}{5} p \sin(2p) + \frac{9}{20} p \sin(3p)$ for spin $j = 3/2$.
\end{enumerate}
For all the above calculations, we put the Immirzi parameter and the Planck length to $1$ for simplifying the notation. 

\section{Summary}
\noindent
Recent results in LQG have discovered that the symmetries of quantum space-time are deformed compared to the gauge structure of general relativity as made explicit in the modification of the HDA. In this work we have shown that, by taking the Minkowski limit of the deformed HDA, one can derive MDRs which are sensitive to several quantization ambiguities through the form of the deformation function. Thus, forthcoming tests of the dispersion relation may allow us to differentiate between different quantization schemes adopted in LQG by using experimental observations. Furthermore, we have outlined a method to accommodate other choices not addressed in this letter, such as the effect of the vacuum state in LQG \cite{LQGvac}, on MDRs. The importance of these choices become paramount in their ramifications for the development of the theory, such as for black-hole entropy calculations (depends on $\gamma$) \cite{bhamb} or for calculating the spectral dimension \cite{dimLQG,dim} (depends on the spin of the representation). 

Further explorations are needed in order to fully understand the nature of these quantum modifications of the HDA. A few steps have been already taken in this direction \cite{kowros,ncLQG,polyLQG}, especially in the attempt to link them to the known structure of Hopf algebras. Here, we have laid a foundation for constructing phenomenological falsifiability conditions for such deformations, dependent on quantization ambiguites within LQG, to be verified by incipient data. We hope this may motivate additional efforts in the QG research community directed both at deriving deformed HDA in other approaches and at investigating the connection between deformations of the HDA and deformations of the Poincar\'e algebra.

\section*{Appendix: rotations as spatial diffeomorphisms}
\noindent
In this section we explicitly show how to derive the bracket between two generators of rotations, i.e. $\{J_l, J_k\}$, from the HDA.  Rotations are generated by the momentum constraint $D[N^{i}]$, since they produce tangential deformations of the hypersurfaces, with shift vector given by $N^i = R^{i}_{l}x^{l} = \epsilon^{ijl}\varphi_{j}x_{l}$ (where $\epsilon^{ijl}$ is the Levi-Civita symbol and $\varphi_{j}$ stands for the angle of a rotation around the $j$ axis). This can be easily understood as follows. Let us introduce a local Cartesian frame on $g_{ij}$ and consider a rotation around the $z$ axis (i.e. we are choosing $j = 3$). Then, the rotated coordinates are obtained just adding $N^i = \epsilon^{i3l}\varphi_{3}x_{l}$ to the starting coordinates $(x,y,z)$. In fact, we have that $x'^i = x^i +N^i$ since in this way we find $x' = x - \varphi_3 y$, $y' = y+ \varphi_3 x$, and $z' = z$, as we could expect. Having proven that $D[N^{i}]$ accounts for rotations, let us derive the Poisson bracket between two Lorentz generators of infinitesimal rotations (i.e. $\{ J_l, J_j \}$) from the hypersurface deformation algebra. In light of the above discussion, this can be done by inserting $N^l = \epsilon^{lik}\varphi_{i1}x_k$ and $M^j = \epsilon^{jmn}\varphi_{m2}x_n$ into
\begin{equation}
\label{hypmom}
\{ D[N^{l}], D[M^{j}] \} = D[\mathcal{L}_{N^i}M^j]
\end{equation}
and, doing so, we obtain
\begin{eqnarray}
\mathcal{L}_{N^i}M^j &=& N^{i}\partial_{i}M^{j}-M^{i}\partial_{i}N^{j}\nonumber\\
&=& \epsilon^{ilk}\varphi_{l1}x_k \epsilon^{jmn}\varphi_{m2}\delta_{ni} -\epsilon^{imn}\varphi_{m2}x_n\epsilon^{jlk}\varphi_{l1}\delta_{ki}\nonumber\\
&=& (\delta_{lj}\delta_{km}-\delta_{lm}\delta_{kj})\varphi_{l1} \varphi_{m2}x_k \nonumber\\
& &\hspace{1cm} -(\delta_{mj}\delta_{nl}-\delta_{ml}\delta_{nj})\varphi_{l1}\varphi_{m2}x_n\nonumber\\
&=& \varphi_{j1}\varphi_{k2}x_k -\varphi_{l1}\varphi_{j2}x_l \nonumber\\
&=& -\epsilon^{jlk}\epsilon_{lts}\varphi_{t1}\varphi_{s2}x_k = -\epsilon^{jlk} \varphi_{l3}x_k
\end{eqnarray}
This means that the right-hand side of Eq. \eqref{hypmom} (i.e. the result of combining two rotations) is still a momentum constraint that implements infinitesimal rotations by an amount $\varphi_{l3}x_k = \epsilon_{lts}\varphi_{t1}\varphi_{s2}x_k $ or, in other words, we have shown that $\{ J_l, J_j \} = \epsilon_{ljk}J_k$.


\section*{Acknowledgements}
\noindent
The authors would like to thank Giovanni Amelino-Camelia, Martin Bojowald, and Antonino Marciano for useful discussions. MR thanks Fudan University of Shanghai for the hospitality during an early stage of the project. MR acknowledges Sapienza University of Rome and MIUR for their partial support through a mobility grant and a starting grant within the Horizon 2020 research program. The contribution of M. R. is based upon work from COST Action MP1405 QSPACE, supported by COST (European Cooperation in Science and Technology).



\end{document}